\documentstyle[12pt]{article}
\begin{document}

\tolerance=5000

\def\pp{{\, \mid \hskip -1.5mm =}}
\def\cL{{\cal L}}
\def\be{\begin{equation}}
\def\ee{\end{equation}}
\def\bea{\begin{eqnarray}}
\def\eea{\end{eqnarray}}
\def\tr{{\rm tr}\, }
\def\nn{\nonumber \\}
\def\e{{\rm e}}
\def\D{{D \hskip -3mm /\,}}

\  \hfill
\begin{minipage}{3.5cm}
April 2003 \\
\end{minipage}

\vfill

\begin{center}
{\large\bf deSitter brane universe 
induced by phantom and quantum effects}

\vfill

{\sc Shin'ichi NOJIRI}\footnote{nojiri@cc.nda.ac.jp}
and {\sc Sergei D. ODINTSOV}$^{\spadesuit}
$\footnote{
odintsov@mail.tomsknet.ru Also at ICREA and IEEC, Barcelona, Spain}

\vfill

{\sl Department of Applied Physics \\
National Defence Academy,
Hashirimizu Yokosuka 239-8686, JAPAN}

\vfill

{\sl $\spadesuit$
Lab. for Fundamental Studies,
Tomsk State Pedagogical University,
634041 Tomsk, RUSSIA}

\vfill

{\bf ABSTRACT}

\end{center}

Five-dimensional braneworld cosmology with deSitter (inflationary) 
brane universe induced by classical and quantum matter is discussed.
It is shown that negative energy phantom field with quantum CFT 
supports the creation of deSitter universe. On the same time,
pure phantom or dust with quantum effects, or Chaplygin gas 
with quantum effects may naturally lead to the occurence of 
Anti-deSitter brane universe but not deSitter one.
It is also interesting that unlike to four-dimensional gravity,
for phantom with (or without) quantum contribution the standard 
cosmological energy conditions may be effectively satisfied.

\vfill

\noindent
PACS: 98.80.Hw,04.50.+h,11.10.Kk,11.10.Wx

\newpage

\noindent
1. The interpretation of recent astrophysical data \cite{SuNv} 
indicate to the acceleration of the scale factor of 
the observable universe. It looks quite possible now that not only
the early universe passed the inflationary stage, but also 
observable universe is (or enters) in deSitter phase.
The origin of deSitter universe could be caused by the dark energy.
There are recent speculations\cite{negative, Gibbons} considering 
negative energy (phantom) theories which produce 
the accelerating scale factor. Unfortunately, such phantom theories 
break the standard cosmological energy conditions. However,
the combination of phantom matter with quantum effects may
improve the situation with energy conditions 
(see, for instance, \cite{nop}).

From another side, the dark energy naturally appears in the
braneworld cosmology (see \cite{dark} and references therein).
It is sometimes easier to achieve brane FRW-cosmology with 
accelerating scale factor even without brane matter.
Then it is quite interesting to consider braneworld cosmology 
with phantom brane matter and quantum brane CFT in order to
check the possibility for deSitter universe occurence.
The present Letter is devoted to the study of above question.
It is shown that 4d brane deSitter occurs for combination
of brane matter as phantom and quantum CFT
or phantom, quantum CFT and dust.
Moreover, in such a case NEC and WEC may be effectively satisfied.
 For pure phantom brane matter
(unlike to four-dimensional case\cite{Gibbons, nop}) there is no
deSitter solution (only Anti-deSitter one) 
but all energy conditions are satisfied.
For combination of matter as Chaplygin gas and quantum CFT
there occurs only brane Anti-deSitter universe.   

\ 

\noindent
2. Let the 3-brane is embedded into the 5d bulk space as in ref.\cite{SMS}.
Let $g_{\mu\nu}$  be the metric tensor of the bulk space and $n_\mu$ be the unit vector 
normal to the 3-brane. Then the metric $q_{\mu\nu}$ induced on the brane has the 
following form: 
\be
\label{S1b}
q_{\mu\nu}=g_{\mu\nu} - n_\mu n_\nu\ .
\ee
The initial gravitational action is
\bea
\label{S00}
S &=&\int d^5 x \sqrt{-g}\left\{ {1 \over \kappa_5^2} R^{(5)} - 2\Lambda + \cdots \right\} \nn
&& + \int \sqrt{-q}\left( - 2\lambda + {\rm matter\ Lagrangian\ density}\right)\ .
\eea
Here $\Lambda$ is the bulk cosmological constant, $\lambda$ is the tension of the brane. 
In the following, the 5d quantities are denoted by the suffix $^(5)$ and 4d ones by 
$^(4)$. In (\ref{S00}), $\cdots$ expresses the matter contribution.  
The bulk Einstein equation is given by
\be
\label{S2bb}
{1 \over \kappa_5^2}\left( R^{(5)}_{\mu\nu} - {1 \over 2}g_{\mu\nu} R^{(5)}\right)
= T_{\mu\nu}\ 
\ee
If one chooses the metric near the brane as:
\be
\label{S2b}
ds^2 = d\chi^2 + q_{\mu\nu} dx^\mu dx^\nu\ ,
\ee
the energy-momentum tensor $T_{\mu\nu}$ has the following form:
\be
\label{S2c}
T_{\mu\nu} = T_{\mu\nu}^{\rm bulk\ matter} - \Lambda g_{\mu\nu} 
+ \delta(\chi)\left(-\lambda q_{\mu\nu} + \tau_{\mu\nu}\right)\ .
\ee
Here $T_{\mu\nu}^{\rm bulk\ matter}$ is the energy-momentum tensor of the bulk matter  
and $\tau_{\mu\nu}$ expresses the contribution due to brane matter. 
Without the bulk matter ($T_{\mu\nu}^{\rm bulk\ matter}=0$),  following the procedure 
in \cite{SMS, KannoSoda}, the bulk Einstein equation can be mapped into the equation on the brane:
\bea
\label{S2d}
&& {1 \over \kappa_5^2} \left( R^{(4)}_{\mu\nu} - {1 \over 2} q_{\mu\nu}R^{(4)}\right) \nn
&& = - {1 \over 2}\left( \Lambda + {\kappa_5^2 \lambda^2 \over 6} \right) q_{\mu\nu} 
+ {\kappa_5^2 \lambda \over 6}\tau_{\mu\nu} 
+ \kappa_5^2\pi_{\mu\nu} - {1 \over \kappa_5^2}E_{\mu\nu}\ .
\eea
Here $\pi_{\mu\nu}$ is given by
\be
\label{S2e}
\pi_{\mu\nu}=-{1 \over 4}\tau_{\mu\alpha}\tau_\nu^{\ \alpha} + {1 \over 12}\tau \tau_{\mu\nu} 
+ {1 \over 8}q_{\mu\nu}\tau_{\alpha\beta} \tau^{\alpha\beta} - {1 \over 24}q_{\mu\nu}\tau^2\ .
\ee
On the other hand, $E_{\mu\nu}$ is defined by the bulk Weyl tensor 
$C^{(5)}_{\mu\nu\rho\sigma}$: 
\be
\label{S6bbb}
E_{\mu\nu}=C^{(5)}_{\alpha\beta\gamma\delta}n^\alpha n^\gamma q_\mu^{\ \beta} 
q_\nu^{\ \delta}\ .
\ee
Note that one may identify the effective 4d gravitational constant 
$\kappa_4$ and 4d cosmological 
constant $\Lambda_4$ with 
\be
\label{S4bb}
{1 \over \kappa_4^2}
={6 \over \lambda \kappa_5^4} \ ,\quad 
\Lambda_4 = {\kappa_5^2 \over 2}\left( \Lambda + {\kappa_5^2 \lambda^2 \over 6} \right) \ .
\ee
As one of the matter fields on the brane, we may consider the phantom field $C$, whose 
energy-momentum tensor has the following form:
\be
\label{PB1}
\tau_{C\, \mu\nu} = \partial_\mu C \partial_\nu C - {1 \over 2}g_{\mu\nu} g^{\alpha\beta}
\partial_\alpha C \partial_\beta C\ .
\ee
This scalar field has negative energy density and its 
solution in 4d deSitter space is \cite{Gibbons}
 
\be
\label{PB2}
C=at + b\ ,
\ee
where a,b are some arbitrary constants.
With above phantom the negative energy density $\rho_C$ and the negative pressure $p_C$  is 
given by
\be
\label{PB3}
\rho_C=p_C=-{a^2 \over 2}
\ee
It was argued in \cite{nop} that phantom behaves as some effective QFT
 in deSitter space. 

The space is assumed to be 5d AdS space, where
\be
\label{EE6}
R^{(5)}_{\mu\nu\rho\sigma}= - {1 \over l^2}\left(g_{\mu\rho} g_{\nu\sigma} - g_{\mu\sigma} 
g_{\nu\rho}\right)\ .
\ee  
Here $\Lambda=-{6 \over \kappa_5^2 l^2}$. Then 
$C^{(5)}_{\alpha\beta\gamma\delta}n^\alpha n^\gamma q_\mu^{\ \beta} q_\nu^{\ \delta}=0$. 
We also assume the brane is the 4d deSitter space, whose metric is taken 
in the following form:
\be
\label{dS1}
ds^2= - dt^2 + L^2 \cosh^2 {t \over L}d\Omega_3^2\ .
\ee
Here $d\Omega_3^2$ is the metric of the 3d sphere with unit radius. 
Then $(t,t)$-component of (\ref{S2d}) has the following form:
\bea
\label{PB4}
0&=& - {3 \over \kappa_5^2 L^2} + {1 \over 2}\left( - {6 \over \kappa_5^2 l^2} 
+ {\kappa_5^2 \lambda^2 \over 6} \right) + {\kappa_5^2 \lambda \over 6} 
\left(-{a^2 \over 2} + \rho_m\right) \nn
&& + \kappa_5^2 \left( {a^4 \over 48} - {a^2 \rho_m \over 12} + {\rho_m^2 \over 12} 
\right)\ ,
\eea
and $(i.j)$-component corresponding to the 3d sphere part is:
\bea
\label{PB5}
0&=& {3 \over \kappa_5^2 L^2} - {1 \over 2}\left(  - {6 \over \kappa_5^2 l^2} 
+ {\kappa_5^2 \lambda^2 \over 6} \right) + {\kappa_5^2 \lambda \over 6} 
\left(-{a^2 \over 2} + p_m\right) \nn
&& + \kappa_5^2 \left( {a^4 \over 16} - {a^2 \rho_m \over 6} - {a^2 p_m \over 12}
+ {\rho_m^2 \over 12} + {\rho_m p_m \over 6}\right)\ .
\eea
Here $\rho_m$ and $p_m$ express the (classical or quantum) 
contribution from the matter on the brane besides the phantom contribution. 
The bulk cosmological constant $\Lambda$ is 
\be
\label{PB6}
\Lambda = - {6 \over \kappa_5^2 l^2} \ .
\ee
By combining (\ref{PB4}) and (\ref{PB6}), one gets
\bea
\label{PB7}
0&=& {\kappa_5^2 \lambda \over 6} 
\left( - a^2 + \rho_m + p_m \right) \nn
&& + \kappa_5^2 \left( {a^4 \over 12} 
 - {a^2 \rho_m \over 4} - {a^2 p_m \over 12} + {\rho_m^2 \over 6}
+ {\rho_m p_m \over 6}\right)\ ,\\
\label{PB8}
0&=& - {6 \over \kappa_5^2 L^2} + \left( - {6 \over \kappa_5^2 l^2} 
+ {\kappa_5^2 \lambda^2 \over 6} \right) + {\kappa_5^2 \lambda \over 6} 
\left( \rho_m - p_m \right) \nn
&& + \kappa_5^2 \left( - {a^4 \over 24} + {a^2 \rho_m \over 12} + {a^2 p_m \over 12}
 - {\rho_m^2 \over 12} - {\rho_m p_m \over 6} \right)\ .
\eea

First we consider the case that there are no matter fields besides the phantom $C$ on the 
brane, that is, $\rho_m=p_m=0$. Then Eq.(\ref{PB7}) gives $a^2=0$, or $a^2=2\lambda$. 
For the latter case, Eq.(\ref{PB8}) gives $0={1 \over L^2} + {1 \over l^2}$.
Then there might be anti-deSitter solution where $L^2=-l^2<0$
 but there is no inflationary brane 
solution (compare with \cite{Gibbons}).
 
As a next step, we may include the quantum effects from the conformal brane matter. 
As usually, the simplest way to do so is to  consider the conformal anomaly: 
\be
\label{OVII}
\tau^A=b\left(F^{(4)}+{2 \over 3} \Box R^{(4)}\right) + b' G^{(4)} 
+ b''\Box R^{(4)}\ ,
\ee
where $F^{(4)}$ is the square of 4d Weyl tensor, $G^{(4)}$ is 
Gauss-Bonnet invariant.
In general, with $N$ scalar, $N_{1/2}$ spinor, $N_1$ vector fields, $N_2$  ($=0$ or $1$) 
gravitons and $N_{\rm HD}$ higher derivative conformal scalars, $b$, $b'$ and $b''$ are 
given by
\bea
\label{bs}
&& b={N +6N_{1/2}+12N_1 + 611 N_2 - 8N_{\rm HD} 
\over 120(4\pi)^2}\nn 
&& b'=-{N+11N_{1/2}+62N_1 + 1411 N_2 -28 N_{\rm HD} 
\over 360(4\pi)^2}\ , \quad b''=0\ .
\eea
Note that $b'$ is negative for the usual matter ($N_{\rm HD}=0$). 

For inflationary brane (\ref{dS1}), one gets
\be
\label{phtm2}
\rho_A=-p_A = - {6b' \over L^4}\ .
\ee
Then Eqs.(\ref{PB7}) and (\ref{PB8}) are: 
\bea
\label{PB11}
0&=& - {\kappa_5^2 a^2 \lambda \over 6} + \kappa_5^2 \left( {a^4 \over 12} 
+ {b' a^2 \over L^4} \right)\ ,\\
\label{PB12}
0&=& - {6 \over \kappa_5^2 L^2} + \left( - {6 \over \kappa_5^2 l^2} 
+ {\kappa_5^2 \lambda^2 \over 6} \right) + {2\kappa_5^2 \lambda b' \over L^4} \nn
&& + \kappa_5^2 \left( - {a^4 \over 24} + {3b'^2 \over L^8} \right)\ .
\eea
Especially when there is no phantom field ($a=0$) and 
$\lambda={6 \over l\kappa_5^2}$,
Eq.(\ref{PB11}) is trivial and Eq.(\ref{PB12}) has the following form:
\be
\label{EE13}
\left({1 \over l} - {\kappa_5^2 b' \over L^4}\right)^2 
= {1 \over L^2}\left(1 + {L^2 \over l^2}\right)\quad \mbox{or}\quad 
\pm {1 \over L}\sqrt{1 + {L^2 \over l^2}} -{1 \over l} = - {\kappa_5^2 b' \over L^4}\ ,
\ee
which (plus sign) reproduces the result  \cite{NO}(see also, \cite{NOZ,HHR}). In other words,
we demonstrated that for the particular choice of the boundary terms, our equation describes 
the quantum creation of deSitter (inflationary) brane which glues two AdS spaces.
Such inflationary brane-world scenario is sometimes called Brane New World \cite{HHR}.

When $a^2\neq 0$, Eq.(\ref{PB11}) gives
\be
\label{PB14}
a^2 = - {12 b' \over L^4} + 2\lambda\ .
\ee
Since usually $b'<0$, if $\lambda>0$, it does not conflict with 
$\lambda={6 \over l\kappa_5^2}$. 
Substituting (\ref{PB14}) into (\ref{PB12}), one gets
\be
\label{PB15}
0=  - {6 \over \kappa_5^2 l^2} - {6 \over \kappa_5^2 L^2} 
 - {\kappa_5^2 \lambda b' \over L^4} - {3\kappa_5^2 b'^2 \over L^8} \ .
\ee
Generally Eq.(\ref{PB15}) has non-trivial solution when $\lambda$ is large enough. 
If we define a new variable $y$ by $y={1 \over \lambda L^4}$, 
Eq.(\ref{PB15}) can be rewritten as 
\be
\label{PB17}
0=\lambda^2 \kappa_5^2 \left\{ - 3{b'}^2 \left(y + {1 \over 6b'}\right) + {1 \over 12} 
\right\} - {6 \over \kappa_5^2 l^2} - {6 \over \kappa_5^2}\sqrt{\lambda y}\ .
\ee
Then in the limit that $\lambda$ is large, one has a deSitter solution 
\be
\label{PB18}
{1 \over L^2}=0\ ,\quad \mbox{or}\quad {1 \over L^2}= \lambda y = - {\lambda \over 3b'}
>0\ .
\ee
Thus, we demonstrated that braneworld which contains phantom 
and quantum CFT on the brane admits deSitter brane solution gluing 
two AdS bulks. It is interesting that in pure 4d space the same type of matter 
also leads to occurence of inflationary universe \cite{nop} in easier way.

 As a next case, we consider the situation that the matter 
on the brane is dust

\be
\label{PB18b}
\rho_d={\alpha \over L^3}\ , \quad p_d=0\ .
\ee
If also the phantom field $C$ presents on the brane, 
Eqs.(\ref{PB7}) and (\ref{PB8}) are: 
\bea
\label{PB19}
0&=& - {\kappa_5^2 a^2 \lambda \over 6}\left( - a^2 + {\alpha \over L^2}\right) 
+ \kappa_5^2 \left( {a^4 \over 12} - {\alpha a^2 \over L^3} 
+ {\alpha^2 \over L^6}\right) \ ,\\
\label{PB20}
0&=& - {6 \over \kappa_5^2 L^2} + \left( - {6 \over \kappa_5^2 l^2} 
+ {\kappa_5^2 \lambda^2 \alpha \over 6L^3} \right) 
+ \kappa_5^2 \left( - {a^4 \over 24} + {a^2 \alpha \over 12 L^3} 
- {\alpha^2 \over 12 L^6}\right)\ .
\eea
Eq.(\ref{PB19}) can be solved as
\be
\label{PB21}
{\alpha \over L^3}={1 \over 2}\left\{ - \left({\lambda \over 6} - a^2\right) 
\pm \sqrt{{2a^4 \over 3} + {\lambda a^2 \over3} + {\lambda^2 \over 36}}\right\}\ .
\ee
Then $+$-sign in (\ref{PB21}) always gives a non-trivial solution if $\lambda>0$. 
If $0<\lambda<{a^2 \over 2}$, 
the $-$-sign in (\ref{PB21}) is also a solution. 

We may consider the combination of the dust and the quantum effect coming from 
the conformal anomaly (\ref{phtm2}), where
\be
\label{DC1}
\rho_m = {\alpha \over L^3}-{6b' \over L^4}\ ,\quad p_m={6b' \over L^4}\ .
\ee
Without phantom field ($a=0$), Eqs.(\ref{PB7}) and (\ref{PB8}) 
are:
\bea
\label{DC2}
0&=& {\kappa_5^2 \alpha \over 6L^3} 
\left( \lambda + {\alpha \over L^3} - {6b' \over L^4} \right)\ ,\\
\label{DC3}
0&=& - {6 \over \kappa_5^2 L^2} + \left( - {6 \over \kappa_5^2 l^2} 
+ {\kappa_5^2 \lambda^2 \over 6} \right) + {\kappa_5^2 \lambda \over 6} 
\left( {\alpha \over L^3} - {12b' \over L^4}\right) \nn
&& - {\kappa_5^2 \over 12}\left( {\alpha \over L^3} - {6b' \over L^4}\right) 
\left( {\alpha \over L^3} + {6b' \over L^4}\right) \ .
\eea
Then in order that Eq.(\ref{DC2}) has non-trivial solution, one gets
\be
\label{DC4}
{6b' \over L^4} = \lambda + {\alpha \over L^3} \ .
\ee
By substituting (\ref{DC4}) into (\ref{DC3}), we obtain
\be
\label{DC5}
{6 \over \kappa_5^2 L^2}=-{6 \over \kappa_5^2 l^2}
 - {\lambda^2\kappa_5^2 \over 12} \ .
\ee
Since the right-hand-side of (\ref{DC5}) is negative, there is no deSitter solution, 
where $L^2>0$ although there might be an anti-deSitter solution, where $L^2<0$. 
Thus, in such situation brane deSitter space does not occur.

\ 

\noindent
3. There are several standard types of the energy conditions
which are usually assumed to be fulfilled in cosmology:
\begin{itemize}
\item Null Energy Condition (NEC): $\rho + p \geq 0$. 
\item Weak Energy Condition (WEC): $\rho\geq 0$ and $\rho + p \geq 0$. 
\item Strong Energy Condition (SEC): $\rho + 3 p \geq 0$ and $\rho + p \geq 0$.
\item Dominant Energy Condition (DEC): $\rho\geq 0$ and $\rho \pm p \geq 0$.  
\end{itemize}
They are violated in four-dimensional space for pure phantom field.
Nevertheless, when phantom field is considered in combination 
with quantum CFT some of the energy conditions survive in 4d deSitter universe\cite{nop}.
Let us check what happens with them in our braneworld.

Eqs.(\ref{PB4}) and (\ref{PB5}) give the effective energy density $\rho_{\rm eff}$ 
and the effective pressure $p_{\rm eff}$ 
\be
\label{PB23}
\rho_{\rm eff}= -{a^2 \over 2} + \rho_m + {6 \over \lambda}
\left( {a^4 \over 48} - {a^2 \rho_m \over 12} + {\rho_m^2 \over 12} \right)\ ,
\ee
and $(i.j)$-component corresponding to the 3d sphere part has the following form:
\be
\label{PB24}
p_{\rm eff}= -{a^2 \over 2} + p_m + {6 \over \lambda} 
\left( {a^4 \over 16} - {a^2 \rho_m \over 6} - {a^2 p_m \over 12}
+ {\rho_m^2 \over 12} + {\rho_m p_m \over 6}\right)\ .
\ee
If there is no non-trivial matter besides the phantom field, one obtains
\be
\label{EE1}
\rho_{\rm eff}= p_{\rm eff}= - {a^2 \over 2} + {a^4 \over 4\lambda}=
{a^2\left(a^2 - 2\lambda\right) \over 4}\ .
\ee
Then if $a^2 > 2\lambda$, 
all the energy conditions are effectively satisfied 
even for pure phantom field. This is an interesting feature
 of phantom braneworld,
as phantom leads to breaking of all energy conditions 
in usual 4d deSitter space\cite{Gibbons}. 
On the other hand, without the phantom field 
\be
\label{EE3}
\rho_{\rm eff}=\rho_m \left(1 + {\rho_m \over 2\lambda}\right)\ ,\quad 
p_{\rm eff}=p_m + {1 \over \lambda}\left({\rho^2 \over 2} + \rho_m p_m\right)\ .
\ee
Then if $0>\lambda > - {\rho \over 2}$, 
 $\rho_{\rm eff}<0$ even if $\rho_m=0$. 
Therefore WEC and DEC are effectively violated. 
Furthermore since
\be
\label{EE5}
\rho_{\rm eff} + p_{\rm eff}=\left(\rho_m + p_m\right)\left(1 + {\rho_m \over \lambda}\right)\ ,
\ee
if $0>\lambda> - \rho_m$,
which is the condition weaker than (\ref{EE5}), all the energy conditions are effectively violated even if 
$\rho_m + p_m \geq 0$ and $\rho>0$.  This is known property of deSitter braneworld.
If we consider the contribution from phantom field and the conformal anomaly, 
Eqs.(\ref{PB23}) and (\ref{PB24}) are
\bea
\label{EE7b}
\rho_{\rm eff} &=& \left( -{a^2 \over 2} - {6b' \over L^4}\right)\left\{1  + {1 \over 2\lambda}
\left( -{a^2 \over 2} - {6b' \over L^4} \right)\right\}\ ,\nn
\label{EE8}
p_{\rm eff}&=& -{a^2 \over 2}  + {6b' \over L^4} + {6 \over \lambda} 
\left( {a^4 \over 16} + {b' a^2 \over 2L^4}  - 3{b'}^2 \right)\ .
\eea
Then 
\be
\label{EE9}
\rho_{\rm eff} + p_{\rm eff} = a^2\left\{- 1 + {1 \over 2\lambda}\left(a^2 + {12 b' \over L^4}\right)\right\}
\ee
When $\lambda>0$, if $a^2<-{12b' \over L^4}$ or $a^2> -{12b' \over L^4} + 4\lambda$, we have $\rho_{\rm eff}>0$ 
and if $a^2>-{12b' \over L^4} - 2\lambda$, we have $\rho_{\rm eff} + p_{\rm eff}>0$. 
On the other hand, when $\lambda<0$, if $-{12b' \over L^4}>a^2> -{12b' \over L^4} + 4\lambda$, 
we have $\rho_{\rm eff}>0$ and if $a^2<-{12b' \over L^4} - 2\lambda$, we have $\rho_{\rm eff} + p_{\rm eff}>0$. 
Then, for example, NEC is effectively satisfied if 
\bea
\label{EE10}
& a^2\geq -{12b' \over L^4} - 2\lambda\quad &(\lambda>0) \nn
& a^2\leq -{12b' \over L^4} - 2\lambda\quad &(\lambda<0) \ ,
\eea
and WEC is effectively satisfied if 
\bea
\label{EE11}
& -{12b' \over L^4} - 2\lambda \leq a^2 \leq -{12b' \over L^4}\ \mbox{or}\ 
a^2\geq -{12b' \over L^4} + 4\lambda \quad &(\lambda>0) \nn
& -{12b' \over L^4}>a^2> -{12b' \over L^4} + 4\lambda \quad &(\lambda<0) \ .
\eea
Thus, in the presence of phantom and quantum CFT it is easier 
to fulfill the standard energy conditions in the braneworld.

\ 

\noindent
4. As a matter, one may consider Chaplygin gas\cite{Gibbons2}, which satisfies
\be
\label{Ch1}
p_{Ch}=-{A \over \rho_{Ch}}\ ,
\ee
where $A$ is a positive constant. For simplicity, we consider $a=0$ case, that is, 
the case without phantom. Then Eqs.(\ref{PB7}) and (\ref{PB8}) have the following form:
\bea
\label{Ch2}
0&=& \left(\rho + \lambda\right)\left(\rho^2 - A\right) \ ,\\
\label{Ch3}
0&=& - {6 \over \kappa_5^2 L^2} + \left( - {6 \over \kappa_5^2 l^2} 
+ {\kappa_5^2 \lambda^2 \over 6} \right) \nn
&& + {\kappa_5^2 \lambda \over 6} \left( \rho_{Ch} + {A \over \rho_{Ch}} \right) 
+ \kappa_5^2 \left(  - {\rho_{Ch}^2 \over 12} + {A \over 6} \right)\ .
\eea
The solutions of (\ref{Ch2}) are easily obtained by
\be
\label{Ch4}
\rho_{Ch} = \pm \sqrt{A}\ ,\quad \rho_{Ch}=-\lambda\ .
\ee
By substituting the former solution $\rho=\pm \sqrt{A}$ into (\ref{Ch3}), one obtains
\be
\label{Ch5}
{6 \over \kappa_5^2 L^2}= {\kappa_5^2 \over 12}\left\{\left(\sqrt{A} 
\mp 2\lambda\right)^2 - 2 \lambda^2 - {72 \over \kappa_5^4 l^2}\right\}\ .
\ee
Therefore in order that $L$ is real, we obtain the following condition
\be
\label{Ch6}
\sqrt{A} < \pm 2\lambda - \sqrt{2\lambda^2 + {72 \over \kappa_5^4 l^2}}\ \mbox{or}\ 
\sqrt{A} > \pm 2\lambda + \sqrt{2\lambda^2 + {72 \over \kappa_5^4 l^2}} \ .
\ee
On the other hand,  choosing the latter solution $\rho = -\lambda$ in (\ref{Ch4}) 
and substituting it into (\ref{Ch3}), one gets
\be
\label{Ch7}
{6 \over \kappa_5^2 L^2} = - {6 \over \kappa_5^2 l^2} - {\kappa_5^2 \lambda^2 \over 12}\ .
\ee
Since the right-hand-side of (\ref{Ch7}) is negative, there is no real solution for $L$. 

We may add the contribution due to the conformal anomaly (\ref{phtm2}) to (\ref{Ch1}): 
\be
\label{CCh1}
\rho_m = \rho_{Ch} - {6b' \over L^4}\ ,\quad 
p_m = {A \over \rho_{Ch}} + {6b' \over L^4}\ .
\ee
Then instead of (\ref{Ch2}) and (\ref{Ch3}), one obtains
\bea
\label{CCh2}
0&=& \left(\rho + \lambda - {6b' \over L^4}\right)\left(\rho^2 - A\right) \ ,\\
\label{CCh3}
0&=& - {6 \over \kappa_5^2 L^2} + \left( - {6 \over \kappa_5^2 l^2} 
+ {\kappa_5^2 \lambda^2 \over 6} \right) + {\kappa_5^2 \lambda \over 6} \left( \rho_{Ch} 
+ {A \over \rho_{Ch}} - {12b' \over L^4} \right) \nn
&& - {\kappa_5^2 \over 12}\left(\rho_{Ch} - {6b' \over L^4}\right)
\left(  \rho_{Ch} - {2A \over \rho_{Ch}} + {6b' \over L^4}\right)\ .
\eea
The solution of (\ref{CCh2}) is:
\be
\label{CCh4}
\rho_{Ch} = \pm \sqrt{A}\ ,\quad \rho_{Ch}=-\lambda + {6b' \over L^4}\ .
\ee
Substituting the former solution $\rho=\pm \sqrt{A}$ into (\ref{Ch3}), we get
\be
\label{CCh5}
{6 \over \kappa_5^2 L^2}= {\kappa_5^2 \over 12}\left\{\left(\sqrt{A} 
\mp \left(2\lambda - {6b' \over L^4}\right)\right)^2 
 - 2 \lambda^2 - {72 \over \kappa_5^4 l^2}\right\}\ .
\ee
It is difficult to solve (\ref{CCh5}) with respect to $L^2$ but 
 in order to get real $L$, the following condition 
should be satisfied:
\bea
\label{CCh6}
\sqrt{A} &<& \pm \left(2\lambda - {6b' \over L^4}\right)
- \sqrt{2\lambda^2 + {72 \over \kappa_5^4 l^2}}\nn
\mbox{or}\ \sqrt{A} &>& \pm \left(2\lambda - {6b' \over L^4}\right)
+ \sqrt{2\lambda^2 + {72 \over \kappa_5^4 l^2}} \ .
\eea
Substituting the latter solution $\rho = -\lambda
+ {6b' \over L^4}$ (\ref{CCh4}) into (\ref{CCh3}), one gets
\be
\label{CCh7}
{6 \over \kappa_5^2 L^2} = - {6 \over \kappa_5^2 l^2} - {\kappa_5^2 \lambda^2 \over 12}\ .
\ee
Since the right-hand-side of (\ref{CCh7}) is negative, there is no real solution 
for $L$. Then there might be an anti-deSitter solution, where $L^2<0$  but there is 
no deSitter solution, where $L^2>0$.

To conclude, the combination of Chaplygin gas with quantum CFT
( unlike to phantom with quantum CFT) 
does not support the occurence of deSitter brane in AdS bulk.
On the same time, the combination of brane quantum CFT and  
phantom may lead to the inducing of 4d deSitter universe. 
It would be interesting to develop such inflationary braneworld 
scenario in more detail.

\ 

\noindent
{\bf Acknowledgments} 
The work by S.N. is supported in part by the Ministry of Education, 
Science, Sports and Culture of Japan under the grant n. 13135208.

\end{document}